\documentstyle[preprint,aps]{revtex}
\tightenlines

\def\singlespace {\smallskipamount=3.75pt plus1pt minus1pt
                  \medskipamount=7.5pt plus2pt minus2pt
                  \bigskipamount=15pt plus4pt minus4pt
                  \normalbaselineskip=12pt plus0pt minus0pt
                  \normallineskip=1pt
                  \normallineskiplimit=0pt
                  \jot=3.75pt
                  {\def\smallskip {\vskip\smallskipamount}}
                  {\def\medskip   {\vskip\medskipamount}}
                  {\def\bigskip   {\vskip\bigskipamount}}
                  {\setbox\strutbox=\hbox{\vrule
                    height10.5pt depth4.5pt width 0pt}}
                  \parskip 7.5pt
                  \normalbaselines}
\def\middlespace {\smallskipamount=5.625pt plus1.5pt minus1.5pt
                  \medskipamount=11.25pt plus3pt minus3pt
                  \bigskipamount=22.5pt plus6pt minus6pt
                  \normalbaselineskip=22.5pt plus0pt minus0pt
                  \normallineskip=1pt
                  \normallineskiplimit=0pt
                  \jot=5.625pt
                  {\def\smallskip {\vskip\smallskipamount}}
                  {\def\medskip   {\vskip\medskipamount}}
                  {\def\bigskip   {\vskip\bigskipamount}}
                  {\setbox\strutbox=\hbox{\vrule
                    height15.75pt depth6.75pt width 0pt}}
                  \parskip 11.25pt
                  \normalbaselines}
\def\doublespace {\smallskipamount=7.5pt plus2pt minus2pt
                  \medskipamount=15pt plus4pt minus4pt
                  \bigskipamount=30pt plus8pt minus8pt
                  \normalbaselineskip=30pt plus0pt minus0pt
                  \normallineskip=2pt
                  \normallineskiplimit=0pt
                  \jot=7.5pt
                  {\def\smallskip {\vskip\smallskipamount}}
                  {\def\medskip   {\vskip\medskipamount}}
                  {\def\bigskip   {\vskip\bigskipamount}}
                  {\setbox\strutbox=\hbox{\vrule
                    height21.0pt depth9.0pt width 0pt}}
                  \parskip 15.0pt
                  \normalbaselines}
\def\al{\alpha}

\def\th{\theta}

\def\si{\sigma}

\def\ph{\phi}

\def\Ph{\Phi}

\def\cF{{\cal F}}
\def\cL{{\cal L}}

\def\frac#1#2{\textstyle{{{#1} \over {#2}}}}

\def\prt{\partial}

\def\half{{\textstyle{1\over 2}}}
\def\lsim{\mathrel{\rlap{\lower4pt\hbox{\hskip1pt$\sim$}}
    \raise1pt\hbox{$<$}}}
\def\gsim{\mathrel{\rlap{\lower4pt\hbox{\hskip1pt$\sim$}}
    \raise1pt\hbox{$>$}}}

\def\etal {{\it et al.}}

\newcommand{\beq}{\begin{equation}}
\newcommand{\eeq}{\end{equation}}
\newcommand{\bea}{\begin{eqnarray}}
\newcommand{\eea}{\end{eqnarray}}

\tightenlines
\begin{document}
\preprint{
\hfill$\vcenter{\hbox{\bf IUHET-480} \hbox{September
             2004}}$  }

\title{\vspace*{.75in}
Lorentz Violation in Supersymmetric Field Theories\footnote{Presented at the
3rd Meeting on CPT and Lorentz Symmetry (CPT 04), Bloomington, Indiana, 
4-7 Aug 2004.}}

\author{M. S. Berger
\footnote{Electronic address:
berger@indiana.edu}}

\address{
Physics Department, Indiana University, Bloomington, IN 47405, USA}

\maketitle

\thispagestyle{empty}

\begin{abstract}
Broken spacetime symmetries might emerge from a fundamental physical 
theory. The effective low-energy theory might be expected to exhibit 
violations of supersymmetry and Lorentz invariance. Some illustrative 
models which combine supersymmetry and Lorentz violation are described, and 
a superspace formulation is given.
\end{abstract}

\newpage

\section{Introduction}

There has been an increasing realization in recent years that the Lorentz 
and Poincar\'e symmetries assumed almost universally in models of particle 
physics might in fact be approximate symmetries that emerge from some
more fundamental theory of quantum gravity. Other potential 
spacetime symmetries such as supersymmetry have yet to be uncovered and 
would have to be broken symmetries
if we are to reconcile them with experimental physics. The interesting 
questions then involve how is the scale of the symmetry breaking determined
in each case. Viewed from the Planck scale, the scale of 
Standard Model symmetry breaking and electroweak-scale supersymmetry are very
small. If the Lorentz symmetry is indeed broken, one of the most pressing 
issues would be to understand why the size of the physical effects are so
incredibly tiny to have escaped all efforts to observe them experimentally. 
There seem to be at least some parallels in the violations of these
spacetimes symmetries, it has motivated some preliminary investigations
to understand any possible connection between them. Of course, since a fully
solvable theory of quantum gravity is not available, the issue can not be 
addressed directly. Rather one must take a more phenomenological approach,
allowing for all possible effects that are consistent with the remaining
symmetries of the theory which might be either exact or spontaneously broken.

The approach taken in Ref.\ \cite{Berger:2001rm} 
is similar in spirit to the Minimal 
Supersymmetric Standard Model (MSSM) where supersymmetry breaking terms are 
added to the Standard Model where all particle fields have been expanded to 
include supermultiplets. Adding terms to a supersymmetric model 
that break the Lorentz symmetry while preserving the supersymmetry can be 
accomplished by modifying (deforming) the supersymmetric algebra and the 
supersymmetric transformation, or less generally one can leave the 
supersymmetric transformation unmodified. In the extensions to the 
Wess-Zumino model described below 
the supersymmetric algebra is modified, so the important 
issues are whether the algebra closes for the supersymmetric transformations
when they are
applied to the fields in the model. The approach is also in the spirit of the
Standard Model Extension (SME)\cite{Colladay:1996iz,Colladay:1998fq}
where Lorentz (and CPT) violating terms are 
introduced into the Standard Model Lagrangian. When one adds the requirement
of supersymmetry, there emerge relationships between the Lorentz-violating
coefficients in a fashion similar to how masses and couplings become related
in a conventional case (MSSM, for example). In 
Refs.\ \cite{Berger:2001rm}
all possible Lorentz-violating terms were added to the Wess-Zumino model which
is a theory involving only a single chiral supermultiplet. These 
simple models do admit a superspace formulation\cite{Berger:2003ay}, and this
motivates future systematic studies in more realistic and interesting 
supersymmetric models.

When supersymmetric particles are discovered at colliders, 
is it possible that Lorentz-violating effects could be 
experimentally interesting? If the effects are as 
suppressed as they appear to be for 
the observed particle content of the Standard Model, then
it will be impossible to observe any new effects. In principle 
the Lorentz-violating effects could arise from terms in 
the Lagrangian involving so far unobserved 
superpartners to the Standard Model particles.
From the point of view of 
phenomenology, this would mean that there are terms in the low-energy 
Lagrangian that violate both supersymmetry and the Lorentz 
symmetry\cite{Belich:2003fa}.
These terms could be less suppressed than the analogous terms in the SME. 
Presumably physical effects will appear radiatively in Standard Model 
physics, and constraints can be derived using existing bounds. 

It is well-known that the assumption 
of Lorentz invariance is needed in quantum field theory to avoid problems with
microcausality. Field theories with Lorentz-violating terms should be
regarded as effective theories and the issues involving microcausality will
be addressed when the full character of the underlying fundamental 
theory emerges at the Planck scale\cite{Kostelecky:2000mm}. 
While the supersymmetric theories described here should be regarded as toy
models, the experimental implications of Lorentz and 
CPT violation parameterized in this manner have been explored extensively in 
recent years\cite{cpt01}.

\section{Superspace}

Lorentz violation has been studied using superfields defined on superspace.
Superspace is defined in terms of spacetime and superspace 
coordinates\cite{Salam:1974yz}
\bea
&&z^M=(x^\mu,\th ^\al,\bar\th _{\dot\al})\;,
\eea
where $\th ^\al$ and $\bar\th _{\dot\al}$ 
each form two-component anticommuting Weyl spinors.
A superfield $\Ph(x,\th,\bar\th)$ is then
a function
of the commuting spacetime coordinates $x^\mu$ and of four anticommuting 
coordinates $\th ^\al$ and $\bar\th _{\dot\al}$.
A chiral superfield is a function of 
$y^\mu =x^\mu+i\th \si ^\mu \bar \th$ and
$\th$. Since the expansion in powers of $\theta$ eventually terminates
this can be expanded as follows
\bea 
\Ph(x,\th,\bar\th)&=&\ph(y)+\sqrt{2}\th \psi(y)+(\th \th)\cF(y)\;, \nonumber \\
&=&\ph(x)+i\th \si ^\mu \bar\th\prt _\mu \ph(x)
-{1\over 4}(\th \th)(\bar\th \bar\th)\Box \ph(x)\nonumber \\
&&+\sqrt{2}\th \psi(x)+i\sqrt{2}\th \si ^\mu \bar\th \th\prt _\mu \psi(x)
+(\th \th)\cF(x)\;.
\eea
The chiral superfield can be described in terms of a differential operator
$U_x$ which is defined as 
\bea
&&U_x \equiv e^{iX}\;, 
\eea
where 
\bea
&&X\equiv (\th \sigma ^\mu \bar\th)\prt _\mu\;. 
\eea
Then an expansion of $U_x$ yields
\bea
&&U_x =1+i(\th \sigma ^\mu \bar\th)\prt _\mu -{1\over 4}(\th \th)
(\bar\th \bar\th)\Box \;. 
\eea
This operator effects a shift $x^\mu\to y^\mu$.
Since the chiral superfield $\Ph(x,\th,\bar\th)$ is a function of $y^\mu$ and 
$\th$ only, the only dependence on $\bar\th$ is in $y^\mu$, 
so it must then be of the form $\Ph(x,\th,\bar\th)=U_x\Psi(x,\th)$
for some function $\Psi$ which depends only on $x^\mu$ and $\th$.

The first supersymmetric model with Lorentz and CPT violation involved 
extending the Wess-Zumino model\cite{Wess:tw}.
The Wess-Zumino Lagrangian can be derived from the superspace 
integral
\bea
\int d^4\th \Ph^*\Ph + \int d^2\th \left [ 
{1\over 2}m\Ph^2 +{1\over 3}g\Ph^3 
+h.c.\right ]\;,
\label{superspace}
\eea
where the conjugate superfield is 
\bea 
\Ph^*(x,\th,\bar\th)&=&\ph^*(z)+\sqrt{2}\bar\th \bar\psi(z)
+(\bar\th \bar\th)\cF^*(z)
\;, 
\eea
where $z^\mu=y^{\mu *}=x^\mu -i\th \si ^\mu \bar \th$.
The superspace integral over $\int d^4\th$ projects out the 
$(\th \th)(\bar \th \bar\th)$ component of the $\Ph^* \Ph$ 
superfield while the $\int d^2\th$ projects out the 
$\th \th$ component of the superpotential.
The result 
\bea
\cL_{WZ}&=&\prt _\mu \phi^* \prt ^\mu \ph 
+{i\over 2}[(\prt _\mu \psi) \si ^\mu 
\bar\psi+(\prt _\mu\bar\psi)\bar\si ^\mu \psi]
+\cF^*\cF \nonumber \\
&&+m\left [\ph \cF + \ph ^*\cF^* -\half\psi\psi -\half\bar\psi\bar\psi\right ] 
\nonumber \\
&&+g\left [\ph^2\cF+\ph^{*2}\cF^*-\ph (\psi\psi)-\ph^*(\bar\psi\bar\psi)
\right ]\;,
\label{wz}
\eea
is a Lagrangian
which transforms into itself plus a total derivative under a supersymmetric 
transformation. The procedure just outlined is well-known and forms a basis
for constructing Lorentz-violating models involving chiral superfields.

\section{Lorentz Violation}

Two Lorentz-violating extensions to the Wess-Zumino model were 
found\cite{Berger:2001rm}, and these two models
admit a superspace formulation\cite{Berger:2003ay}.
Define new operators that can act on superfields as 
\bea
&&U_y \equiv e^{iY}\;, \\
&&T_k \equiv e^{-K}\;.
\eea
where 
\bea
&&Y\equiv k_{\mu\nu}(\th \sigma ^\mu \bar\th)\prt ^\nu\;, \\
&&K\equiv k_\mu(\th \sigma ^\mu \bar\th)\;.
\eea
The expansions are 
\bea
&&U_y =1+ik_{\mu\nu}(\th \sigma ^\mu \bar\th)\prt^\nu
 -{1\over 4}k_{\mu \nu}k^{\mu \rho}(\th \th)
(\bar\th \bar\th)\prt ^\nu \prt _\rho\;, \\
&&T_k =1-k_\mu(\th \sigma ^\mu \bar\th)+{k^2\over 4}(\th \th)
(\bar\th \bar\th)\;.
\eea
Here $k_{\mu\nu}$ and $k_\mu$ are Lorentz-violating coefficients that 
transform under observer Lorentz transformations but do not transform 
(or transform as a scalar) under
particle Lorentz transformations. They therefore represent possible 
descriptions of physically relevant effects.
Since $Y$, like $X$, is a derivative operator, the action of $U_y$
on a superfield ${\mathcal S}$
is a coordinate shift. The appearance of terms of 
order ${\mathcal O}(k^2)$ in the Lagrangians 
is easily understood in both cases in terms of 
these operators.
Furthermore we have $U_y^*=U_y^{-1}$ 
while $T_k^*=T_k$ and not its inverse.

The supersymmetric models with Lorentz-violating terms 
can be expressed in terms of new superfields,
\bea
\Ph_y(x,\th,\bar\th)&=&U_yU_x\Psi(x,\th)\;, \\
\Ph^*_y(x,\th,\bar\th)
&=&U_y^{-1}U_x^{-1}\Psi^*(x,\bar\th)\;.
\eea
Applying $U_y$ to the chiral and antichiral superfields merely effects the
substitution $\prt _\mu \to \prt _\mu +k_{\mu\nu}\prt ^\nu$. Since $U_y$ 
involves a derivative operator just as $U_x$, the derivation of the chiral 
superfield $\Ph_y$ is a function of the variables 
$x_+^\mu=x^\mu + i\th \si ^\mu \bar\th + ik^{\mu \nu}\th\si_\nu\bar\th$
and $\theta$ analogous to how, in the conventional case, $\Ph$ is a function 
of the variables $y^\mu$ and $\theta$. The 
Lagrangian is given by 
\bea
&&\int d^4\th \Ph_y^*\Ph_y+ \int d^2\th \left [ 
{1\over 2}m\Ph_y^2 +{1\over 3}g\Ph_y^3 
+h.c.\right ]\nonumber \\
&&=\int d^4\th \left [U_y^*\Ph^*\right ]
\left [U_y\Ph\right ]+ \int d^2\th \left [ 
{1\over 2}m\Ph^2 +{1\over 3}g\Ph^3 
+h.c.\right ]\;.
\label{superspace2}
\eea

For the CPT-violating model the superfields have the form
\bea
\Ph_k(x,\th,\bar\th)&=&T_kU_x\Psi(x,\th)\;, \\
\Ph^*_k(x,\th,\bar\th) 
&=&T_kU_x^{-1}\Psi^*(x,\bar\th)\;.
\eea
It is helpful to note that the transformation $U_x$ acts on $\Psi$ and 
its inverse $U_x^{-1}$ 
acts on $\Psi^*$, while the same transformation $T_k$ acts on both 
$\Psi$ and $\Psi^*$ (since $T_k^*=T_k$). A consequence of this fact is 
that the supersymmetry transformation will act differently on the components 
of the chiral superfield and its conjugate. 
Specifically the chiral superfield $\Ph_k$ is the
same as $\Ph$ with the substitution $\prt _\mu \to \prt _\mu +ik_\mu$ 
whereas the 
antichiral superfield $\Ph_k^*$ is the same as $\Ph^*$ with the 
substitution $\prt _\mu \to \prt _\mu -ik_\mu$.

The CPT-violating model 
can then be represented in the following way as a 
superspace integral:
\bea
&&\int d^4\th \Ph_k^*\Ph_k=\int d^4\th \Ph^* e^{-2K}\Ph
\label{proj}
\eea
Unlike the CPT-conserving model, the $(\th \th)(\bar\th \bar\th)$ component
of $\Ph^*\Ph$ no longer transforms into a total derivative. A specific  
combination of components of $\Ph^*\Ph$ does transform into a total 
derivative, and this combination is in fact the 
$(\th \th)(\bar\th \bar\th)$ component of $\Ph_k^*\Ph_k$.

The Lagrangians for the two models in terms of the component fields 
can be found in Refs.\ \cite{Berger:2001rm,Berger:2003ay}.

\section{Conclusions}

Lorentz-violating extensions of supersymmetric theories
model can be understood 
in terms of analogous transformations on  
modified superfields and projections arising from superspace integrals.
Such superspace formulations should allow efficient investigations into 
possible Lorentz violation in more complicated theories.


\section*{Acknowledgments}
This work was supported in part by the U.S.
Department of Energy under Grant No.~DE-FG02-91ER40661.


\begin{thebibliography}{xx}
\def\etal {{\it et al.}}

\bibitem{Berger:2001rm}
M.~S.~Berger and V.~A.~Kostelecky,
{\it Phys.\ Rev.\ } {\bf D65}, 091701 (2002).

\bibitem{Colladay:1996iz}
D.~Colladay and V.~A.~Kostelecky,
{\it Phys.\ Rev.\ } {\bf D55}, 6760 (1997).

\bibitem{Colladay:1998fq}
D.~Colladay and V.~A.~Kostelecky,
{\it Phys.\ Rev.\ } {\bf D58}, 116002 (1998).

\bibitem{Berger:2003ay}
M.~S.~Berger,
{\it Phys.\ Rev.\ } {\bf D68}, 115005 (2003).

\bibitem{Belich:2003fa}
H.~Belich, J.~L.~Boldo, L.~P.~Colatto, J.~A.~Helayel-Neto and A.~L.~Nogueira,
{\it Phys.\ Rev.\ } {\bf D65}, 065030 (2003).

\bibitem{Kostelecky:2000mm}
V.~A.~Kostelecky and R.~Lehnert,
{\it Phys.\ Rev.\ } {\bf D63}, 065008 (2001).

\bibitem{cpt01}
See, for example,
V.A.\ Kosteleck\'y, ed.,
{\it CPT and Lorentz Symmetry II},
World Scientific, Singapore, 2002.

\bibitem{Salam:1974yz}
A.~Salam and J.~Strathdee,
{\it Nucl.\ Phys.\ } {\bf B76}, 477 (1974).

\bibitem{Wess:tw}
J.~Wess and B.~Zumino,
{\it Nucl.\ Phys.\ } {\bf B70}, 39 (1974).

\end{thebibliography}
\end{document}